\documentclass[sigconf]{acmart}

\renewcommand\footnotetextcopyrightpermission[1]{}
\acmConference[CHIWORK '26 Workshop]{%
 CHIWORK '26 Workshop: Interrogating GenAI Augmentation for CHIworkers}{%
 June 22, 2026}{Linz, Austria}
\acmYear{2026}

\usepackage{graphicx}
\usepackage{float}

\newcommand{\workshopNote}{%
  \vspace{-0.22em}
  \begin{quote}
    \small\itshape
    This position paper was presented at the CHIWORK '26 Workshop
    \emph{Interrogating GenAI Augmentation for CHIworkers:
    Strategies for Professional Autonomy and Accountability}
    (June 22, 2026, Linz, Austria). Workshop proposal:
    \cite{sandhaus2026interrogating}.
  \end{quote}%
}

\title{From Human-Centered Design to Human-AI Collaboration: Why the Future of HCI Still Starts With People}

\author{Issam Alzouby}
\affiliation{
  \institution{University of North Carolina at Charlotte}
  \city{Charlotte}
  \country{USA}}
\email{ialzouby@charlotte.edu}

\begin{document}

\begin{abstract}
Generative AI is reshaping HCI work by accelerating brainstorming, writing, prototyping, and interface generation. However, faster production does not automatically produce human-centered design. This position paper argues that GenAI should be treated not as a replacement for human-centered methods, but as a co-thinking partner driven by human goals, ethics, cognition, and accountability. After reflecting on HCI coursework, design activities, and AI-assisted workflows, I argue that traditional HCD principles become more important as AI systems become more capable. Users still bring limited attention, mental models, trust issues, and cognitive biases into interaction, while GenAI introduces new risks around over-reliance, de-skilling, shallow reasoning, hallucination, unclear authorship, and reduced accountability. The future of HCI should therefore focus on human-AI collaboration that preserves human agency, critical thinking, and responsibility.
\end{abstract}


\maketitle

\workshopNote

\section{Introduction}

In the early days of computing, leaders like Steve Jobs emphasized the idea that technology must feel intuitive and focused on users. A computer was no longer supposed to be a complex engineering machine that was only understood by experts. It was supposed to feel approachable, clean, and understandable. That approach heavily influenced the growth of human-centered design.

Traditional human-centered design was created because engineers realized something important: building powerful systems means nothing if real people cannot use them. Over time, human-centered interaction researchers began prioritizing cognition, perception, motor control, usability heuristics, and contextual inquiry to best understand how humans actually interact with technology.\cite{hci, heuristics,contextual}

Today, computing is entering another major transformation. Instead of users simply clicking buttons and navigating through menus and drop-downs, they are now collaborating with AI systems that can generate content, reason through tasks, summarize information, and even go out and perform the clicking and navigation autonomously. The graphical user interface itself is transforming. A text box connected to a large language model can suddenly replace layers of menus, buttons, and flows.

At first, GenAI felt like the greatest tool for human-centered design. Users could theoretically interact with systems through natural language instead of memorizing systems. Interfaces can adapt dynamically to different users based on their preferences and mental models. But after working through HCI coursework, I do not think GenAI replaces the need for HCD at all. In many ways, it makes HCD even more important.

The rapid growth and adoption of generative AI is also starting to reshape discussions around labor, productivity, and the future role of human expertise. A recent report from Anthropic suggested that AI exposure is becoming increasingly dominant in knowledge-based professions like programming, finance, customer support, and admin work.\cite{anthropicjobs} While replacement has not fully occurred, companies are increasingly using AI systems to augment or partially automate workflows.

Users will no longer be operating systems manually, they will be working as collaborators, or orchestrators. AI systems are becoming participants within workflows rather than a tool on a browser.

\begin{figure}[H]
\centering
\includegraphics[width=\linewidth]{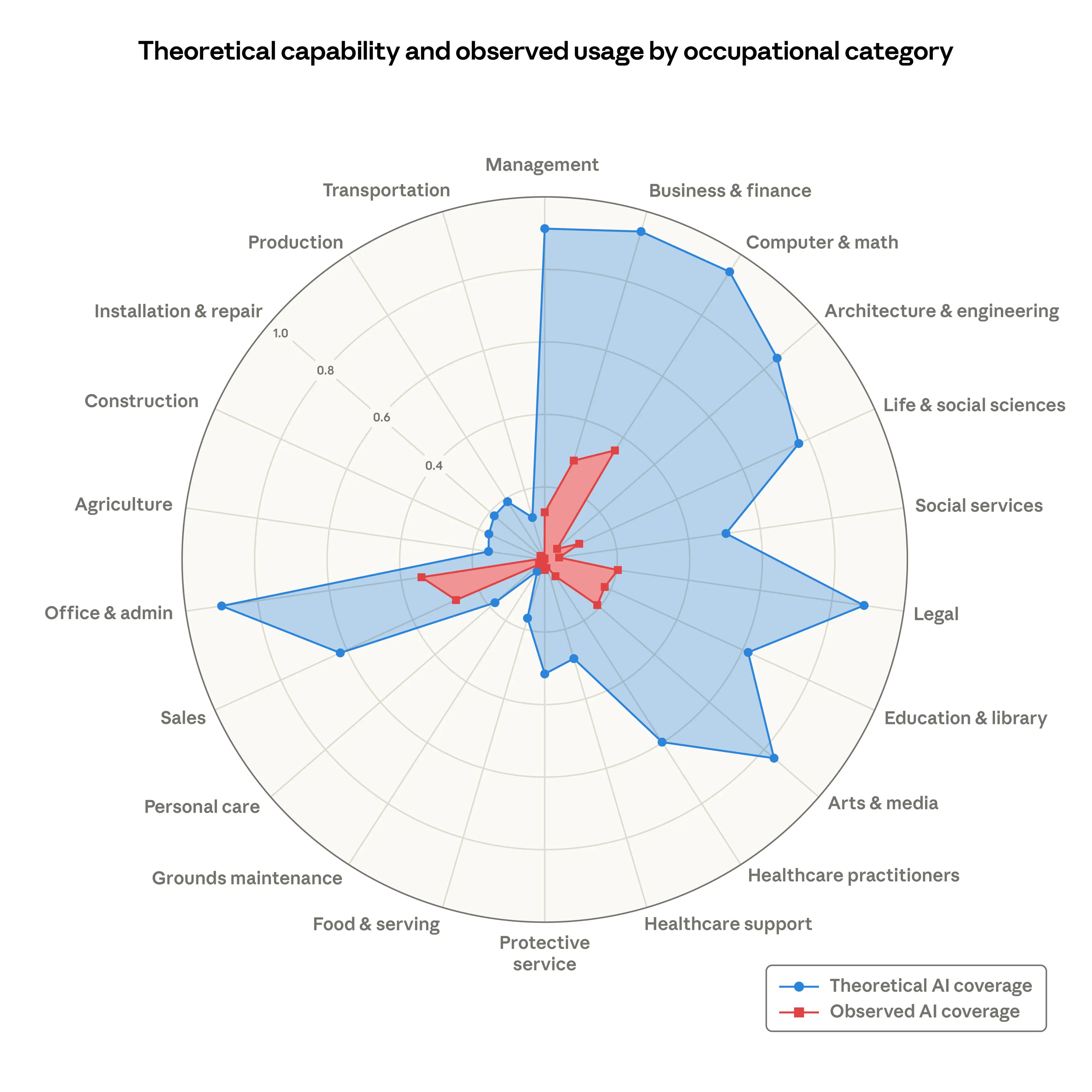}
\caption{Anthropic's analysis of occupations with high AI exposure highlights how knowledge-based and white-collar professions are increasingly affected by generative AI systems.}
\Description{A radar chart comparing theoretical AI capability and observed AI usage across occupational categories, with higher exposure shown in knowledge-based categories such as office administration, computer and math, business and finance, and legal work.}
\label{fig:ai_jobs}
\end{figure}

The central argument of this paper is simple: traditional HCD remains essential because we still have the same limitations, mental models, and trust issues we have always had, which can even vary by person. GenAI will change the user interface in a significant way, but it's not going to resolve human limitations. If anything, AI systems amplify many of the existing HCI concerns while introducing entirely new ones.

\section{The Core Strengths of Traditional HCD}

One of the strengths of traditional HCD is its power in recognizing human faults in an effort to perfect the experience. We as humans have limited attention, limited working memory, and constantly rely on mental models to understand systems.

This sounds obvious now, but back then it was a big change in computing. This shift from Dot-Com-era static HTML webpages to dynamic interactive systems was significant. 

Early research in HCI focused heavily on user cognition, perception, and motor control because interfaces were becoming too complicated for users to naturally understand. Users are not just looking at pixels on a screen. They're developing internal understandings of how they believe the system works.\cite{hci}

That idea became very clear during a semester-long HCI design project. While we performed a low-fidelity prototype, we assumed certain element modifications would feel obvious to users. They did not. In fact, some introduced more confusion. Some participants saw the features completely differently than what we initially intended. Other users ignored sections we thought were visually recognizable. Watching the users actually walk through the system during cognitive walkthroughs like the 'Wizard-of-Oz' helped me understand the difference between real user behavior and what developers imagine as they build the systems.

That is exactly why contextual design became so important in HCI. Rather than designing systems in isolation, contextual design requires the study of users in the environment where the interaction actually happens. The product no longer becomes the primary focus. They are treated as real people operating within constraints, habits, distractions, and workflows.

Another major strength of HCD is iterative evaluation. Nielsen's heuristic evaluation approach showed that usability problems become visible quickly when different evaluators test an interface. That approach still matters today. During the same class project, small observations kept changing big parts of our design. Some of the best improvements came from seeing users hesitate for just a short period of time during interactions.\cite{heuristics}

Traditional HCD also succeeds because it understands that interaction actually matters. Interaction is not just clicking buttons. It involves qualities like explainability, transparency, effectiveness, and how intuitive the system is. Good interfaces reduce cognitive load and confusion.

The ``Code Bubbles'' reading was another great example of this approach. Rather than forcing programmers to constantly navigate confusing and overloading file structures, the interface adapted around how developers naturally think about the active working files and the relationships between code sections. The goal was not just to improve functionality; it was to reduce users’ cognitive load. \cite{codebubbles}

The importance of reducing cognitive load is even more important in AI-assisted systems. Modern users already face overwhelming amounts of information output, redundant notifications, and confusing recommendations. Good HCD doesn't just make systems functional, it also reduces the unnecessary mental effort and lets users focus their attention and cognition on meaningful tasks rather than the interface.

That philosophy is what made a lot of classic tech products feel magical and revolutionary during the early consumer computing era. The success behind Apple's Apple II was based solely on the idea that anyone, in any household, can use these systems. Command lines disappeared into the background and users could focus on goals rather than mechanics.

\section{The Limitations and Challenges of Traditional HCD}

It's important to mention that traditional human-centered design has always struggled with scale, speed, and complicated systems.

One issue is that human-centered tasks are usually slow and expensive, increasing development costs and sometimes even runtime. Contextual inquiry, usability studies, iterative prototyping, and field observations require significant time, effort, and financial investments. In the fast-moving world of big tech, teams tend to skip these stages entirely because they want rapid development.

Another issue is that HCD tends to assume that most systems are stable. Modern systems are no longer static. Applications now update continuously, personalize interfaces dynamically, and process massive amounts of data in real time.

Traditional HCD also struggles when users can't clearly communicate what they actually need. This too became quite clear during our project interviews. Some participants described frustrations with systems they used every day, but couldn't explain exactly why those systems felt frustrating. People often adapt to bad interfaces over time and overlook the quirks that cause friction.

Also, HCD sometimes overemphasizes usability while underemphasizing adoption realities and shareholder expectations. This was something we learned about during our visiting Google guest speaker Aakanksha Parameshwar’s talk.  A product can be usable and still fail commercially or organizationally.\cite{adoption}

Another challenge is that today's systems involve invisible computations. Traditional HCD was largely developed around interfaces where all system controls were visible. AI systems drastically break that model. Today, users do not really see how decisions are being made inside the black box.

\section{How Generative AI Changes HCD}

GenAI changes HCD in two major ways. 

First, it changes how systems are designed.

Second, it changes how users interact with systems.

From a design perspective, GenAI massively accelerates ideation and prototype development. During our project work, AI tools made it dramatically faster to generate interface ideas, rewrite content, brainstorm layouts, and iterate on workflows. At one point, one of our team members even experimented with an AI code generator, which was able to output an entire dynamic version of Expedia, with our changes. This introduces changes such as going directly from idea to high-fidelity without any low-fidelity prototyping required. Tasks that once took hours could now happen in seconds.

But the deeper, more drastic change is the actual interaction itself.

Traditional interfaces require users to adapt to software structures. Users memorized menus, commands, and navigation patterns of their favorite systems. GenAI flips that relationship completely. Now the system adapts to what the human desires most.

Instead of navigating nested menus, users can simply describe goals through natural text, like having a conversation with a colleague. This is aligned with human cognition and also helps reduce memory load. In many ways, GenAI supports an experience that HCI researchers have been chasing for decades.

However, GenAI also introduces entirely new usability problems.

One major issue I have observed through our course readings is interpretability. Research on interpretability tools showed that even data scientists are prone to misunderstanding AI explanation tools and over-trust model outputs. That observation stuck with me because it highlights something even more concerning: humans naturally assume systems are more intelligent and completely reliable. \cite{interpretability}

This becomes especially concerning when interfaces appear conversational and human-like.

Users develop incorrect mental models of AI systems. That problem showed up repeatedly across our class discussions. People assume the model ``understands'' context, reasoning, or truth at a deeper level than it actually does.

GenAI also increases the risk of cognitive offloading. If systems automatically generate writing, decisions, summaries, and handle creativity, users may forget how to think. I recall reading a paper from the MIT Media Lab, warning that heavy reliance on AI tools can contribute to what is described as “cognitive debt,” where users show reduced neural cognition, weaker memory, and lower levels of critical thinking during writing and problem-solving tasks. \cite{mit} This is very concerning because excessive dependence on GenAI could gradually weaken deeper reasoning, learning, and reflective thinking abilities.

Another issue within this scenario is trust.

Traditional HCD focused heavily on predictability and consistency. AI systems are non-deterministic and rely on one-to-many statistics. The same exact prompt can produce different outputs on the same model. These GenAI systems are also infamous for their tendency to hallucinate. Interfaces might appear polished while being completely unfactual.

That creates a strange concern. The interaction feels natural and enables trust, but the true reliability is far less stable.

The cryptocurrency seed phrase paper written by Farida et al. actually connects surprisingly well to this concern. The work found that a lot of users developed incorrect mental models around crypto seed-phrase security systems and the associated recovery mechanisms. Users assumed that cryptocurrency systems worked like traditional banking systems. This shows a classic HCD problem: users interpret new technologies from existing mental models. \cite{seedphrases}

The same thing is happening with AI today.

People tend to treat GenAI outputs as ground truth rather than probabilistic outputs.

\section{Project Experiences and Why HCD Matters More With AI}

The project work throughout the HCI course helped shape how I think about human-centered design and AI systems. At first, I understood user-centered design as a general principle: build systems around users instead of forcing users to adapt to systems. Through interviews, cognitive walkthroughs, low-fidelity prototyping, and iterative revisions, I began to understand the actual process needed to make that principle meaningful.

One thing that stood out during testing was how differently users interpreted the same interface. Some users focused on the visual hierarchy. Others cared more about the clarity of the system flow. Some ignored sections we thought were central to the design. These observations reinforced a key idea from HCI: users do not experience interfaces objectively. Their expectations, prior experiences, mental models, and strategies shape how they interact with a system. \cite{hci}

We also used GenAI tools during parts of the design process. AI was useful for quickly generating interface ideas, rewriting content, brainstorming layouts, and producing prototype code. However, the outputs often lacked contextual awareness. Some designs looked polished but violated basic usability principles. Other features sounded impressive but did not make sense in realistic user flows. This showed me that GenAI can accelerate design work, but it cannot replace the human-centered reasoning needed to evaluate whether a design actually supports users.

This is why I think HCD becomes more important, not less important, as AI systems become more capable. Conversational AI interfaces feel natural because people are used to communicating through language. But that natural interaction style can also create risky assumptions. Users may treat AI systems as expert collaborators rather than probabilistic systems that can produce incomplete, misleading, or hallucinated outputs.

The course readings on interpretability reinforced this concern. Even experienced data scientists can misunderstand AI explanation tools and over-trust model outputs. This creates a major HCI challenge because trust becomes harder to design for. If experts can misread AI explanations, everyday users are even more likely to build incorrect mental models of how these systems work. \cite{interpretability}

Another issue is cognitive offloading. If users rely on AI systems to generate writing, summaries, decisions, and creative ideas too often, they may engage less deeply in independent reasoning. Recent work on AI-assisted writing describes this concern as “cognitive debt,” where heavy reliance on AI tools may reduce memory, reflection, and critical thinking during writing tasks. \cite{mit}

For that reason, I see future HCD practitioners taking on a stronger evaluator role. They may spend less time manually creating every interface element from scratch and more time validating AI-generated outputs against real human needs, workflows, and mental models. In this future, HCD is not replaced by GenAI. It becomes the discipline that keeps AI systems grounded in human behavior, human limitations, and human accountability.

\section{The Future: Human-AI Collaboration, Not Replacement}

I think the future of AI involves integration into our daily lives through the principles within HCD. 

GenAI is powerful, but that is not enough to create good, long-lasting interactions. We have already seen this historically. Early software prioritized technical capability over usability. HCD became a thing because people realized that technical capabilities don't mean much if humans cannot understand or trust the system. Early computers were incredibly powerful, but computing didn't transform our societies until each individual was able to pick up a computer and use it!

I believe that future HCD practitioners should adopt and incorporate GenAI in many ways.

First, AI should be an assistant, or support tool, not something to replace human creativity.

Second, AI models should become more transparent and expand the black box. Users need accurate mental models of model behavior instead of confident outputs. These models need to be explainable not only to experienced data scientists, but also to the average person.

Third, interfaces should encourage human agency. AI should support human decision-making, not replace it. These systems could easily ask you what you think and suggest your own insights to keep you in the loop.

Fourth, HCD practitioners should design for relief of cognitive friction. These systems should help users think better, not think less.

Finally, I think ethical design becomes even more of an issue in an AI-driven world. Bias, misinformation, over-automation, and manipulation can easily become rampant with the use of GenAI systems. If human-centered safeguards are ignored, the consequences could be severe.

I see a future where we do not even rely on physical keyboards and mice, instead controlling systems through speech, gestures, and even thoughts.

The more intelligent these systems become, the more important it becomes to truly understand and build around people.

\section{Conclusion}

Traditional HCD transformed the world of technology by emphasizing a focus on people rather than machines. It introduced approaches and science to ground system development in cognition, perception, usability, and contextual understanding which I believe is still a pillar of effective HCD. \cite{heuristics,contextual,hci}

Generative AI does not eliminate the need for those principles. It makes these principles more important and more crucial to our success.

AI systems are completely changing how users interact with technology, but humans still have the same cognitive limitations, biases, mental models, and trust issues that HCI researchers have studied for a long period of time. If anything, I believe GenAI makes these issues harder to identify because interactions feel more natural and human-like, when in reality, it is the exact opposite.

Throughout the HCI course, I came to see that good design comes from understanding human behavior rather than focusing on faster, more engaging systems. This was true whether we were conducting the heuristic evaluations, contextual inquiries, cognitive walkthrough, or prototyping. The strongest and most impactful systems appear when humanity is at the center of the development process.

The future of user development will not only rely on the most advanced AI models. It will belong to the systems that combine AI capability with deep human understanding in the most effective way.

That was true during the days of Steve Jobs. It is still true today.

\begin{acks}
\textbf{AI Use Statement:} Generative AI tools were used minimally during the preparation of this paper for light language refinement and formatting support. The argument, structure, examples, interpretation, and final content were written, reviewed, and approved by the author.
\end{acks}

\bibliographystyle{ACM-Reference-Format}
\bibliography{references}

\end{document}